\documentclass{mem}
\usepackage{natbib}\usepackage{txfonts}\usepackage{balance}
\usepackage{graphicx}
\usepackage[a4paper,breaklinks,dvipdfm]{hyperref}
\idline{75}{282}
\begin{document}
\def\teff{$T\rm_{eff }$}
\def\Thbb{$T\rm_{HBB }$}
\def\ocen{$\omega\,{\rm {Cen }}$}
\def\kms{$\mathrm {km s}^{-1}$}
\newcommand{\msun}{\ensuremath{\, {M}_\odot}}
\newcommand{\Msun}{\ensuremath{\, {M}_\odot}}

\title{
Lithium in the context of Multiple Populations in Globular Clusters
}


\author{
F. \,D'Antona
          }

\institute{
Istituto Nazionale di Astrofisica --
Osservatorio Astronomico di Roma, Via di Frascati 33,
I-00078 Monteporzio Catone, Italy
\email{franca.dantona@gmail.com}
}

\authorrunning{D'Antona }

\titlerunning{Lithium evolution in GCs}

\abstract{
Multiple Populations represent the standard for Globular Clusters (GC): a fraction (10--50\%) of their stars have the same elemental abundances of halo stars of similar metallicity (first generation, or 1G), but the other stars (second generation, 2G) are characterised by patterns of light elements abundances which resemble those typical of gas processed by  proton--capture reactions at high temperature. Consequently, we should naively expect that Lithium is destroyed in the 2G stars, but instead it is generally observed, at abundances only slightly depleted with respect to the 1G stars.  After discussing the models for the formation of multiple populations, I examine the role of dilution with pristine gas and the possible role of the Asymptotic Giant Branch (AGB) scenario in accounting for the Lithium patterns in GC stars. Super--AGB and AGB yields of Lithium, produced by the Cameron Fowler mechanism in the `Hot Bottom Burning' convective envelopes, may help to explain the peculiar high Lithium in a few extreme 2G stars. On the other hand, modeling the abundances in mild 2G stars depends explicitly on whether the initial Li in the gas forming the 1G stars and the diluting gas of the 2G ones is that predicted by the Big Bang nucleosynthesis or the $\sim$3 times smaller value observed at the surface of halo dwarfs.

\keywords{Stars: abundances -- Stars: Population II -- Galaxy: globular clusters -- 
Galaxy: abundances -- Cosmology: observations }
}
\maketitle{}

\section{Introduction}

\begin{figure*}[t!]
\vskip -80pt
\begin{center}
\includegraphics[scale=0.45]{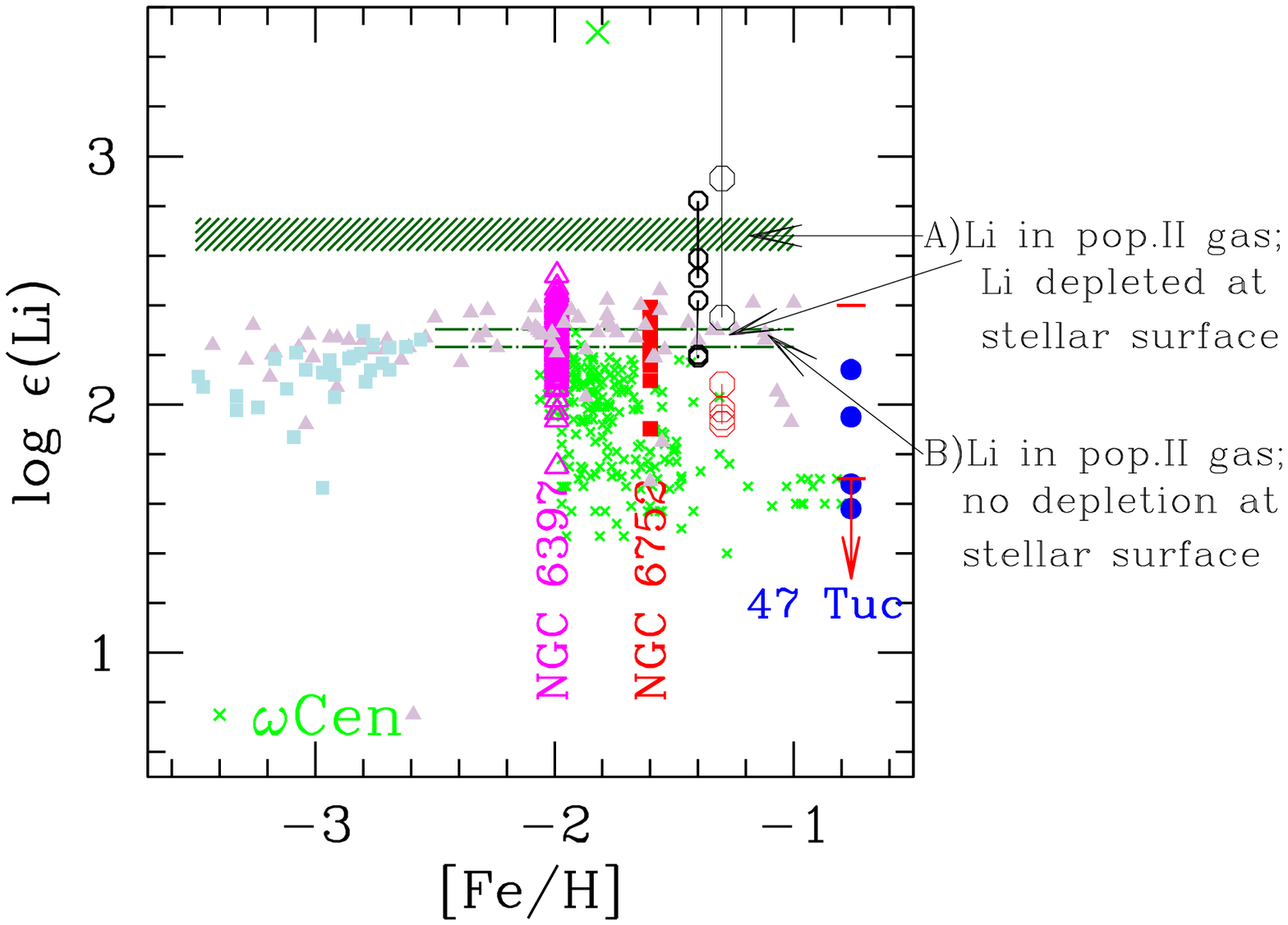}
\caption{\footnotesize
Compilation of data and model results discussed in this talk. In the plane $ log \epsilon$(Li) vs. [Fe/H] we plot some reference data for the Globular Clusters: NGC\,6397 from \cite{lind2009}, NGC\,6752 from \cite{shen2010}, 47\,Tuc from \cite{bonifacio2007} plus the limits by \cite{dorazi2010}, included in the interval limited by the red line and upper limit. The red giant data from \ocen\ by \cite{mucciarelli2018} are plotted as small green crosses, with a big green cross indicating the location of the peculiar giant \#25664 \citep{mucciarelli2019}. The \ocen\ points are scaled up by 1\,dex to include the approximate dilution factor at the first dredge up. The halo subdwarfs data are from \cite{melendez2010} (analysis by non-LTE models, grey triangles) and \cite{sbordone2010} (cyan squares). At [Fe/H]=--1.3 and --1.4 we plot the abundances in the ejecta of AGB models (4--6\Msun: red circles) and super-AGB models (6.5--7.5\Msun: black light circles, the 7.5\msun\ is out of scale) by \cite{venturadantona2010} and the super--AGB yields by \cite{dicrisci2018, dantona2019} (6.5-7.5\Msun, heavy black circles).
The dashed green upper band represents the Big--Bang Lithium abundance, while the lower dash--dotted lines represent the average Li at the surface of halo stars. The two possible interpretations for the initial lithium in the gas forming the 1G GC stars, and for the gas diluting the ejecta in the 2G stars are shown as case A and B. }
\label{eta}
\end{center}
\end{figure*}

There are two scientific problems which are interconnected, when we study the Lithium patterns in the multiple populations (MPops) of GCs.\\
The first one is of special interest to this meeting on  `Lithium in the Universe' and it is: how can we solve the problem of the discrepancy between the Li abundance predicted by the standard Big--Bang cosmology \citep[e.g.][]{pitrou2018} and the observations of the Li abundance at the surface of warm population II stars, which is a factor 2--3 smaller (see Fig.\,\ref{eta} and its references)? Many researchers agree that the reason behind this discrepancy is `simply' due to stellar depletion mechanisms acting at the surface of these stars, and this point of view is supported by the fact that the lowest metallicity stars ([Fe/H]$<$--2.5)\footnote{$\log \epsilon$(Li)=log[n(Li)/n(H)]+12} indeed show smaller Li, with a larger star--to--star spread, than the remarkably similar value $\log \epsilon$(Li)$\sim 2.2-2.3$\ of the warm dwarfs at  $-2.5<$ [Fe/H] $<-1.5$, suggesting a variation in the effect of the depletion mechanisms due to the physical changes in the structure of the lowest metallicity stars. In addition, models by \cite{richard2005}\footnote{It is a pleasure for me to recall that the basics --for pop.\,I-- of these models including diffusion, turbulence and meridional circulation were presented by Georges Michaud in Monteporzio at the workshop ``The problem of Lithium" in 1990 \citep{Michaud1991}}, are plausibly in agreement both with the plateau and with the depletion of Li at [Fe/H]$<$--2.5. \\
The second problem is: how were the GC stars showing composition `anomalies'  formed, and is Lithium a key tool to choose a formation model?

\section{Multiple populations}
GCs display a a great number of signatures that reveal the presence of multiple populations (MPops).
While one population (in general a minority of the cluster stars) has a chemical tagging similar to field halo stars, and therefore is called ``first generation" (1G), the other stars have  chemical abundances of ``light" elements  typical of gas processed at high temperature by proton capture reactions (including $^4$He) and are globally dubbed ``second generation" or 2G. The signatures are found by many different observation tools: 

1) high dispersion spectroscopy reveals that the abundance patterns of light elements display ``anticorrelations", the most relevant of which is the Na vs. O anticorrelation. Anticorrelated are also Mg and Al and Si and Mg, in the (more restricted) number of clusters in which Mg variations are clearly detected \citep[see for a summary ][]{gratton2019}; 

2) traditional photometry in optical or near IR bands has shown the presence of multiple main sequences  in a few clusters, indicating populations with different helium abundance \cite{bedin2004, piotto2007};

3) HST UV spectrophotometry showed that the UV bands F275W and F336W powerfully magnify N (and C and O) abundance differences in the spectra   \citep{piotto2015}. Thus ``chromosome maps", diagrams in two pseudo-colors, were built in which the 1G and different groups of  2G occupy different regions \citep{milone2017}. Chromosome maps allowed to extend to large stellar samples the results of high dispersion spectroscopy, and clearly showed the ubiquity, variety and discreteness of MPops \citep{renzini2015}. For a comparison between high dispersion results and chromosome maps see \cite{marino2019}.

\section{The AGB scenario}
If a model has to deal with the observed abundance patterns, it must, first of all, identify a (stellar) source of gas processed by p--captures at high or very high temperature and then explain how the formation of 2G stars occurs in this gas. We refer the reader to the review by \cite{gratton2019} for a description of the models, and here we limit ourselves to say that the MPops, in GCs where Mg is depleted, are only compatible with the ``supermassive" stars (SMS) scenario \citep{denissenkov2014}, maybe in the excretion-belt version by \cite{gieles2018}, and the AGB scenario \citep{dercole2008}. 
In this latter model, the p--captures occur at the bottom of the very hot convective envelopes of luminous massive AGBs (``hot bottom burning", HBB), the winds 
lost by these stars at low velocity collect in a cooling flow at the center of the GC, where they can form 2G stars, either from the `pure' ejecta, or from the ejecta diluted with re--accreted pristine gas. The AGB scenario has been explored both from the nucleosynthesis of AGBs point of view \citep{ventura2009,ventura2011,ventura2013, ventura2018mgisotopes}, from the dynamical point of view \citep{dercole2008,dercole2016,bekki2011,calura2019} and for its consequences on the long term evolution of the spatial distribution of the two stellar generations \citep[e.g.][]{vesperini2010,vesperini2011,vesperini2013}

The SMS scenario would be favoured as it predicts more correctly the Mg isotopes ratios in the 2G, but \cite{ventura2018mgisotopes} have shown that a reasonable change in the cross section of $^{25}$Mg+p and $^{26}$Mg+p rates in the temperature region of interest ($  100-120$MK) can produce agreement also for the AGB scenario. On the other hand, the SMS scenario has the drawback that no SMS has ever been observed, and it does not explain why Mg depletion is more significant in metal poor GCs, a peculiarity well explained in the AGB scenario \citep{ventura2011}.\\
Is Lithium the key element to choose between the two models?

\section{Lithium patterns and the MPops}
The role of Lithium is indeed very simple. If high temperature p-captures are needed to explain the abundance patterns, these same p-captures have already destroyed $^7$Li at a much lower temperature inside the evolving star! Nevertheless, $^7$Li  is present in 2G stars, as shown by many observations \citep[for a complete summary, see ][]{gratton2019}.

This conundrum has two possible tentative solutions:

1) the gas from which the 2G stars were born is diluted with standard pop.\,II gas, which has preserved the starting abundance. In fact, the typical O--Na, Mg--Al anticorrelations are explained by dilution of ejecta (O-poor, Na-rich, Mg-poor, Al-rich...) with standard pop.\,II gas. Most clusters display a Na--Li anticorrelation too, which can be the signature of dilution of 2G matter, Na rich but Li-free, with matter having standard Na and Li \citep[e.g.][]{decressin2007b}.
On the other hand, \cite{gratton2019} stress that the dilution factors inferred from the observations and the Li in 2G stars require that the ejecta can not be Li--poor.

2) In the AGB scenario, the same HBB which is responsible for the anomalous composition of the 2G stars, also produces fresh lithium by the \cite{cameronfowler1971} mechanism. Li remains at very high abundances in the envelope until $^3$He is totally consumed, and eventually it is fully burned. Thus the massive AGBs can expel Li with the gas lost during their initial HBB phase, and contribute to the Li abundance of the mixture.

A problem with this hypothesis is that the Li average abundance in the ejecta is scarcely constrained, due to its strong dependence on the mass-loss rate during the lithium rich phase \citep[see the still actual discussion in][]{ventura2005b}.
\begin{figure*}[]
\begin{center}
\includegraphics[scale=0.35]{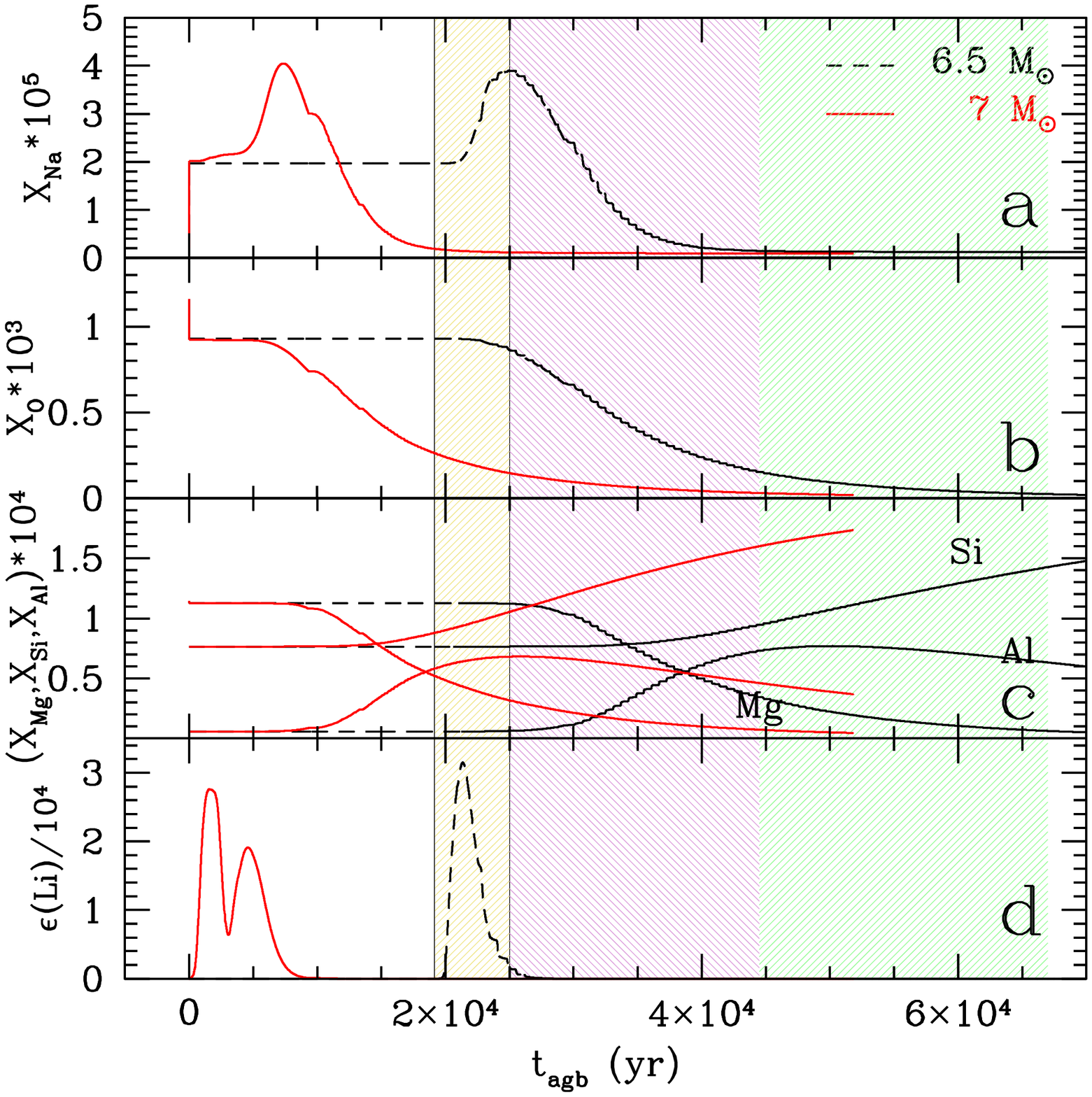} 
\caption{
\footnotesize{Surface abundances of Li, O, Mg and Na along the evolution of super--AGBs of masses 6.5, and 7\,M$_\odot$, from \cite{dicrisci2018,dantona2019}. Different regions of HBB re highlighted for the 6.5\Msun: 1st (yellow): region of Li--production by the Cameron-Fowler mechanism and Na production by burning of the Ne obtained by the second dredge up; 2nd (pink): Na destruction, O and Mg depletion, Al formation; 3rd (green): Si production and steady decrease in Al. } }
\label{timeagb}
\end{center}
\end{figure*}

\section{The peculiar Li--rich 2G stars: in NGC\,2808 and \ocen}
Figure\,\ref{timeagb} shows the time evolution of light elements in the envelopes of super--AGB stars of 6.5 and 7\Msun \citep{dicrisci2018}. Along the evolution, the stellar luminosity increases and the HBB temperature \Thbb\ increases too, favoring more extreme p-captures. So, Li is formed as soon as the star climbs along the AGB, even before the second dredge up is finished (\Thbb$\sim$40\,MK), and also Na reaches its largest abundances, due to the second dredge up and the p-captures on the dredged up $^{22}$Ne. Later on (when \Thbb$>$65\,MK) the NO cycle becomes efficient and Oxygen declines,  while the Mg-Al chain becomes efficient too. In the latest phases, Si builds up (and Al begins declining). The elemental yields therefore come from very different physical conditions (\Thbb) and Li and extreme p-processed gas can coexist. 

The AGB scenario also predicts that the ``extreme'' population  ---the group whose helium abundance is at the highest values--- present in some GCs was formed from pure AGB/super--AGB gas \citep{dercole2008, dantona2016}. 
Thus the elemental abundance in these stars should reflect directly the composition of the ejecta. 

Among several giants observed by \cite{dorazi2015} in the GC NGC\,2808, one (\#46518) belongs to the ``extreme" population and its Li is a bare factor of two smaller than in the 1G giants of the same sample. Note that \cite{dorazi2015} consider in the same class all the giants having a similarly high Al abundance, but actually we have seen in Fig.\,\ref{timeagb} that Al can be reduced when Si is produced, so Al is a very general indicator of 2G, and not an indicator of the most extreme stars only. On the contrary, the location of \#46518 in the chromosome map shows that it has indeed the most extreme composition \citep{marino2019}. 
In \cite{dantona2019} we show that the abundances in the \#46518 giant are compatible with its direct birth of super-AGB ejecta, while its Li abundance requires a large degree of dilution with pristine, Li-rich, gas, if this star was born in the SMS scenario.

An extreme case is that of the \ocen\ giant with extreme Li and Na abundances discovered by \cite{mucciarelli2019}. The star is shown in Fig.\,\ref{eta} as the green cross at the top. It may be compatible with the abundances in the most extreme super-AGB ejecta computed by \cite{venturadantona2010}, and easily explained by looking at Fig.\,\ref{timeagb}: in fact {\it the maximum abundance of Li and Na in the envelopes of super--AGBs (and massive AGBs) occur at the same time}. Therefore, if mass loss occurs mainly at this epoch, as in the most massive super--AGBs, we will find very high Li and Na in the ejecta. Note that a measure of other abundances in these giants may disprove this interpretation: we {\it do not} expect extremely low O and Mg, nor high Al in this star. 

Thus, when the Li observed abundance is very large, in the context of the AGB scenario, we should find a {\it direct} correlation between Li and Na abundance. In all other cases, we expect to find an anticorrelation (as found), which can be attributed to dilution.

\section{Which Li in pristine diluting gas?}

A deceiving point which can be overlooked when considering the role of Li abundance in the ejecta is related to how we interpret the abundances at the surface of pop.\,II dwarfs. This is illustrated in Fig.\,\ref{eta}, where the Li Big Bang abundance \citep[see, e.g.][]{pitrou2018} is shown  by the dashed green band at the top, and the average dwarf location is within the two dash-dotted lines. We can take two different points of view:

(A) the abundance in the gas forming the 1G is the Big Bang abundance, and stellar depletion mechanisms bring it down to the value at the surface of dwarfs;

(B) the abundance in the gas forming the 1G is that seen at the surface of dwarfs.

If the ejecta have no Li it is unimportant if the lower abundance at the surface is already in the gas or it is achieved later on at the stellar surface, and case (A) or (B) give the same result for any mixture. But if the diluting gas has a finite Li abundance as in the AGB ejecta, and similar processes at the stellar surface deplete Li in both 1G and 2G stars, in case (A) the gas abundance resulting from a mixture of pristine and AGB gas will result in a {\it smaller} abundance than in case (B). 
{\it If we want to obtain a particular Li abundance at the surface of a 2G dwarf, in case (A), the Li abundance in the ejecta, for a given degree of dilution with pristine gas, must be a factor $\sim$2.5 larger than in case (B).}
In the simplest exaample,  a mild 2G dwarf, for which we have estimated  a dilution factor is $\sim$0.5, shows a standard pop.II abundance $\log \epsilon$(Li)=2.3. If we want to obtain this abundance in case B, we need  that both the 50\% pristine gas and the 50\% AGB gas have $\log \epsilon$(Li)=2.3. But in case A we need that both have $\log \epsilon$(Li)=2.7 (then depleted at the surface). Thus Li in the ejecta of all AGB forming the 2G should be close to the Big Bang abundance, values which are found in today's models only from the high tail of super--AGB masses (see Fig.\,\ref{eta}). 
A plausible solution to this problem is achieved if the dilution factors for 2G stars (e.g. given in \cite{gratton2019}) have been underestimated.

\section{Conclusions}
The Lithium observations in GC stars are a powerful tool to investigate their MPops, and the abundances in the 2G stars may favour or help to dismiss the different formation models. We find that only exceptional extreme stars of the 2G, such as the candidates discussed here in NGC\,2808 and \ocen\ directly favour the AGB scenario, while the mild 2G stars having only mildly depleted Li abundances may be consistent with a scenario in which the ejecta (of any model) are strongly diluted. \\  
Nevertheless, if the AGB scenario will be recognized as the best model for MPops, we will get a totally independent hint of the reliability of the Big Bang abundance. 



\begin{acknowledgements}
I am grateful to Paolo Ventura and to the other SOC and LOC members for the effort spent in organizing, scientifically and practically, such an exciting meeting, 29 years after the (much smaller) workshop on ``The Problem of Lithium" \citep{dantona1991MmSAI}. Many exciting topics found an assessment from that time, several still remain unsettled, many new problems arise, but a lot of work is still due before we have full solutions to our questions.    
\end{acknowledgements}

\end{document}